\documentclass[%
 reprint,
superscriptaddress,
 amsmath,amssymb,
 aps,
]{revtex4-2}
\usepackage{tikz-feynhand}
\usepackage[autostyle]{csquotes}
\usepackage{float}
\usetikzlibrary{arrows}
\usepackage{graphicx}
\usepackage{dcolumn}
\usepackage{bm}
\usepackage{hyperref}
\usepackage{amsmath}
\usepackage{slashed}
\usepackage{subfigure}
\begin{document}

\preprint{}

\title{$B\to X_c\ell\bar{\nu}$ decay as a probe of complex conjugate poles}

\author{Jinglong Zhu}
\email{Contact author: zhujl23@mails.jlu.edu.cn}
\affiliation{Center for Theoretical Physics and College of Physics, Jilin University, Changchun, 130012, China}
\author{Hiroyuki Umeeda}%
\email{Contact author: umeeda@jlu.edu.cn}
\affiliation{Center for Theoretical Physics and College of Physics, Jilin University, Changchun, 130012, China}
\affiliation{China Center of Advanced Science and Technology, Beijing 100190, China}
\affiliation{Institute of High Energy Physics, Chinese Academy of Sciences,
Beijing 100049, China}

\date{\today}

\begin{abstract}
In this work, nontrivial analytic structure of the quark propagator is discussed for \textit{B}-meson inclusive decays. Attributed to invalidity of the standard K\"{a}ll\'{e}n-Lehman spectral representation, complex conjugate poles alter the evaluation of decay rates, which lead to violation of quark-hadron duality. As phenomenological observables, widths in $B\to X_c\ell\bar{\nu}$ decay as well as the lifetime of the $B_d^0$ meson are discussed. In the presence of the mentioned nonanalytical contributions, the possibility for resolving the $|V_{cb}|$ puzzle is addressed. It is demonstrated that there exists a parameter region, which explains $|V_{cb}|$ and $B_d^0$-meson lifetime simultaneously within 1$\sigma$ via the complex conjugate poles from a charm quark.
\end{abstract}

\maketitle

\section{INTRODUCTION}
Inclusive decays of \textit{B} mesons play a crucial role for the determination of the Cabibbo-Kobayashi-Maskawa (CKM) matrix elements~\cite{Cabibbo:1963yz,Kobayashi:1973fv}. The measurement of decay spectra enables us to extract $|V_{cb}|$ with $1.1\%$-$1.5\%$ precision in Refs.~\cite{Bordone:2021oof, Bernlochner:2022ucr,Finauri:2023kte}. An alternative determination of $|V_{cb}|$ can be made via the exclusive decays such as $B\to D^{(*)}\ell\bar{\nu}$, where \textit{ab initio} simulation of quantum chromodynamics (QCD) is performed by lattice gauge theory \cite{FlavourLatticeAveragingGroupFLAG:2021npn}. Provided that nonperturbative aspects of QCD are properly controlled, the different methods to determine CKM matrix elements supposedly agree with one another.
\par
In this context, it should be noted that there has been a longstanding tension for more than a decade: the discrepancy between inclusive and exclusive $|V_{cb}|$ is observed, where Particle Data Group (PDG) 2024 \cite{ParticleDataGroup:2024cfk} reports significance of 3.0$\sigma$. The interpretation via new physics is disfavored \cite{Crivellin:2014zpa} since this scenario is incompatible with the $Zb\bar{b}$ coupling constraint, while the \textit{BABAR} experiment recently gave the consistent result between both determinations albeit a large uncertainty \cite{BaBar:2023kug}. Precise determinations of $V_{cb}$ improve the accuracy of $\epsilon_K$, representing indirect \textit{CP} violation for $K^0-\bar{K}^0$ mixing.
\par
As to the theoretical framework, a conventional method for inclusive \textit{B}-meson decays is the operator product expansion (OPE) \cite{Wilson:1969zs}, rendering the observables recast into the form of the $1/m_b$ series \cite{Chay:1990da, Bigi:1993fe, Manohar:1993qn, Blok:1993va}. Given that nonanalytic behavior of the charm quark propagator exists solely along the timelike real axis, the perturbative expansion leads to convergent prediction for a contour that is away from the resonance region \cite{Chay:1990da}. Equipped with this methodology, $|V_{cb}|$, bottom quark mass and power-correction parameters are determined so as to fit the decay spectra \cite{Bordone:2021oof, Bernlochner:2022ucr,Finauri:2023kte}.
\par
There is a tacit assumption made in the OPE analysis, called quark-hadron duality \cite{Bloom:1970xb,Bloom:1971ye,Poggio:1975af}. The underlying difficulty arises from truncated power series, i.e., power and radiative corrections, which would lead to particularly nontrivial behavior when analytic continuation to Minkowski spacetime is implemented \cite{Shifman:2000jv}. Violation of quark-hadron duality is hard to quantify and instead modeled by the instanton-based approach \cite{Chay:1994si, Chay:1994dk, Falk:1995yc, Chibisov:1996wf, Umeeda:2022dpt} as well as the resonance-based approach \cite{Colangelo:1997ni, Grinstein:1997xk,Bigi:1998kc,Grinstein:1998gc,Bigi:1999fi,Bigi:1999qe,Lebed:2000gm,Grinstein:2001zq,Grinstein:2001nu,Mondejar:2006ct} in previous works for heavy quarks. It is indicated that while duality violation is exponentially suppressed in Euclidean space, it exhibits oscillatory behavior in Minkowski spacetime \cite{Shifman:2000jv}. For semileptonic $B$-meson decays, global duality violation involving the smearing over phase space might give rise to a potential difficulty in the phenomenological discussion.
\par
Meanwhile, there exists a nontrivial issue on strong interaction: colored particles cannot be observed individually, referred to as color confinement, where its underlying mechanism is still not clarified. In relation to this phenomenon, there are some notable aspects in analytical structures of propagators for confined particles. In particular, the existence of complex conjugate poles (CCPs) is possibly an indication of confinement, invalidating the standard K\"{a}ll\'{e}n-Lehman spectral representation \cite{Umezawa:1951rp, Kallen:1952zz, Lehmann:1954xi}. 
In previous works, CCPs of gluon \cite{Binosi:2019ecz} and quark propagators \cite{Zhu:2020bwu} are extracted by (a variant of) the Schlessinger point method \cite{Schlessinger:1968vsk} supplemented by Euclidean-input methods. From field-theoretic perspective, it has been known \cite{Zwanziger:1989mf} that CCPs for the gluon propagator are suggested in view of the Gribov problem \cite{Gribov:1977wm}. See also recent intensive investigations of CCPs mostly oriented for the Yang-Mills theories \cite{Hayashi:2018giz, Hayashi:2020few, Hayashi:2021nnj, Hayashi:2021jju}.
\par
In this work, contributions of CCPs to inclusive widths for $B$-meson decays are discussed. It should be noted that CCPs lead to novel nonanalytical behavior of the forward scattering tensor in $B\to X_c\ell\bar{\nu}$ decays. In the presence of CCPs, deformation of the integration contour yields corrections that cannot be evaluated perturbatively. These additional contributions can be extracted by the residue theorem, in general invalidating quark-hadron duality. Formulating in this way, we discuss the possibility that the aforementioned $|V_{cb}|$ puzzle is resolved. Furthermore, the contributions of CCPs to nonleptonic $B_d^0$-meson decays are also studied. We shall show that $|V_{cb}|$ and the $B_d^0$ lifetime can be simultaneously explained within $1\sigma$, with certain values of the pole position and the residue.
\par
\section{$B$-MESON SEMILEPTONIC DECAYS}
In the previous works \cite{Chay:1990da, Bigi:1993fe, Manohar:1993qn, Blok:1993va}, inclusive decays of beauty hadrons are evaluated in the heavy quark effective theory (HQET) \cite{Georgi:1990um, Eichten:1989zv, Eichten:1990vp, Mannel:1991mc}, for which the lowest order contribution corresponds to the free bottom quark decay. Here we consider semileptonic decays with massless leptons denoted with $\ell=e, \mu$ at the rest frame of the $B$ meson. The triple differential rate is given by \cite{Manohar:1993qn}
\begin{eqnarray}
     \frac{d^3\Gamma}{dE_{\bar{\nu}}dE_\ell dq^2}=\frac{G_F^2|V_{cb}|^2}{4\pi^3}L_{\mu\nu}W^{\mu\nu},\label{eq1}
\end{eqnarray}
where $L_{\mu\nu}$ and $W^{\mu\nu}$ represent leptonic and hadronic tensors, respectively, with $q=p_{\ell}+p_{\bar{\nu}}$. The latter quantity is related to the imaginary part of the forward scattering tensor, $W^{\mu\nu}=-\frac{1}{\pi}\mathrm{Im}T^{\mu\nu}$, with $T^{\mu\nu}$ defined by
\begin{eqnarray}
     T^{\mu\nu}&=&-i\int d^4xe^{-iq\cdot x} \left\langle H_v |T[J^\mu(x)J^\nu(0)]|H_v\right\rangle
      \label{eq:threea}.
\end{eqnarray}
In Eq.~(\ref{eq:threea}), the hadronic state is normalized via $\left\langle H_{v'}(\bm{p})|H_v(\bm{q})\right\rangle=v^0(2\pi)^3\delta^{(3)}(\bm{p}-\bm{q})$. By introducing a variable of $\alpha=q^2/[2m_b(q^0-E_\ell)]$ and integrating Eq.~(\ref{eq1}) over $q^0$, one can obtain the double differential width
\begin{eqnarray}
     \frac{d^2\Gamma}{dE_\ell d\alpha}&=&
     -\frac{G_F^2|V_{cb}|^2}{4\pi^3}\frac{2m_b}{\pi}
     \nonumber\\
     &\times&\textrm{Im}\int_{E_\ell}^{q^0_\textrm{max}}
     (q^0-E_\ell)L_{\mu\nu}T^{\mu\nu}dq^0.
     \label{Eq:DoubleDiff}
\end{eqnarray}
\begin{figure}[t]
\begin{center}      
\begin{tikzpicture} [scale=0.7]
\begin{feynhand}
    \draw (0, 1.4) circle [radius=0.3cm];
    \draw (0,-1.4) circle [radius=0.3cm];
    \vertex (a1) at (-5,0);
    \vertex (a2) at (5,0);
    \propag [plain] (a1) to (a2);
    \vertex (q1) at (-3,-4);
    \vertex (q2) at (-3,4);
    \propag [->] (q1) to (q2);
    \vertex [particle] (a3) at (-0.5,0);
    \setlength{\feynhandlinesize}{1pt};
    \propag [bos] (a1) to (a3);
    \vertex [particle] (a4) at (3,0);
    \propag [bos] (a2) to (a4);
    \vertex (w1) at (1.5,-3);
    \vertex (w2) at (1.5,0);
    \propag[->] (w1) to (w2);
    \vertex (w3) at (1.5,3);
    \propag  (w3) to (w2);
    \vertex (e1) at (-0.6,3);
    \vertex (e2) at (-0.6,-3);
    \propag [->] (w3) to (e1);
    \propag (w1) to (e2);
    \vertex (e3) at (-2.6,3);
    \vertex (e4) at (-2.6,-3);
    \propag (e1) to (e3);
    \propag [->] (e4) to (e2);
    \vertex (l1) at (-2.6,-0.3);
    \vertex (l2) at (-1.6,-0.3);
    \vertex (l3) at (-0.6,-0.3);
    \propag (l1) to(e4);
    \propag (l2) to (l1);
    \propag [->](l3) to (l2);
    \vertex (u1) at (-2.6,0.3);
    \vertex (u2) at (-1.6,0.3);
    \vertex (u3) at (-0.6,0.3);
    \propag (u1) to(e3);
    \propag [->](u1) to (u2);
    \propag (u2) to (u3);
    \vertex (c1) at (-0.3,0);
    \propag [->](u3) to [in=90, out=0,looseness=0.8] (c1);
    \propag (c1) to [in=0, out=-90,looseness=0.9] (l3);
    \setlength{\feynhanddotsize}{0.7mm};
    \vertex[dot] (m1) at (0,1.4){};
    \vertex[dot] (m2) at (0,-1.4){};
    \vertex (f1) at (-3,-3) ;
    \vertex (f2) at (-3,3) ;
    \vertex (o1) at (-3.5,-3) {$-\infty$};
    \vertex (o2) at (-3.5,3) {$\infty$};
    \propag[sca] (f1) to (e4);
    \propag[sca] (f2) to (e3);
    \vertex (1) at(3.5,3.8);
    \vertex (2) at(3.5,3.2);
    \vertex (3) at(4,3.2);
    \propag (1) to (2);
    \propag (2) to (3);
    \vertex (4) at (3.8,3.5){$q^0$};
    \node at (-1.6,0.6) {$C_a$};
    \node at (-0.5,3.3) {$C_b$};
    \node at (0.5,1.7) {$C_p$};
    \node at (0.5,-1.85) {$C_{p^\prime}$};
    \vertex (p1) at (0.0, 1.7);
    \vertex (p2) at (-0.09, 1.7);
    \propag[->] (p1) to (p2);
    \vertex (p3) at (0.1, 1.1);
    \vertex (p4) at (0.09, 1.1);
    \propag[->] (p4) to (p3);
    \vertex (p5) at (0.05, -1.7);
    \vertex (p6) at (-0.04, -1.7);
    \propag[->] (p6) to (p5);
    \vertex (p7) at (0.0, -1.1);
    \vertex (p8) at (-0.09, -1.1);
    \propag[->] (p7) to (p8);
\end{feynhand}
\end{tikzpicture}
\caption{Integration contours in the complex plane of $q^0=E_\ell+E_{\bar{\nu}}$ in the presence of one-pair CCPs. The wavy line that $C_a$ wraps around is a physical cut for $B\to X_c\ell \bar{\nu}$ while one located on the right side is the one for $b\to bb\bar{c}$ processes at the partonic level.}
\label{fig3}
\end{center}
\end{figure}
The expression of $T^{\mu\nu}$ relevant for $B\to X_c\ell\bar{\nu}$ is
\begin{eqnarray}
\left.T^{\mu\nu}\right|_{B\to X_c\ell\bar{\nu}}
      &=&-i\int d^4xe^{-iq\cdot x}\nonumber\\
      &\times&\left\langle H_v |\bar{b}(x)\gamma^\mu P_L S(x,0)\gamma^\nu P_L b(0)|H_v\right\rangle.\qquad
      \label{eq:three}
\end{eqnarray}
In Eq.~(\ref{eq:three}), $S(x,y)$ represents the charm quark propagator. In the presence of one pair of CCPs, the Green function for the nonperturbative region includes
\begin{eqnarray}
    \left. S(x,y)\right|_{\rm CCPs} &=& \int \frac{d^4p_c}{(2\pi)^4}i\slashed{p}_c\bigg(\frac{R}{p_c^2+Q}+\frac{R^*}{p_c^2+Q^*}\bigg)
    \nonumber\\
    &\times& e^{-ip_c\cdot(x-y)},\label{eq:two}
\end{eqnarray}
where the complex numbers denoted as \textit{R} and \textit{Q} represent the residue and the pole position, respectively. Although there are additional CCP terms that have trivial Dirac structure, we omitted those in Eq.~(\ref{eq:two}) since they vanish in Eq.~(\ref{eq:three}) due to chirality-projection operators. The case with multiple-pair CCPs is considered in the later discussion.
\par
The forward scattering tensor in Eq.~(\ref{eq:three}) can be divided into separate nonanalytical structures
\begin{eqnarray}
T^{\mu\nu}=\tilde{T}^{\mu\nu}+T^{\mu\nu}_{\rm CCP}+T^{\mu\nu}_{\mathrm{CCP}^\prime}.
\label{Eq:Tmunusep}
\end{eqnarray}
In Eq.~(\ref{Eq:Tmunusep}), $\tilde{T}$ represents a contribution that possesses a physical cut along the timelike real axis, which is conventionally analyzed by the perturbation theory, while the other two terms are associated with the CCPs in Eq.~(\ref{eq:two}).
\par
It should be noted that the CCPs do not contribute to the result as long as the integration range is defined by the one in Eq.~(\ref{Eq:DoubleDiff}) since the two terms in Eq.~(\ref{eq:two}) give a real-valued combination to the integrand. However, this is not the case when nontrivial deformation of integration contour is implemented, as explicitly discussed later.
\par
The integration defined on the rhs~of Eq.~(\ref{Eq:DoubleDiff}) can be carried out as follows: the phase space integral range defined by $E_\ell \leq q^0 \leq q^0_{\rm max}$ is deformed in such a way that the contour wraps around the branch cut for $B\to X_c \ell\bar{\nu}$ decays. After this is implemented, the contour is given by $C_a$ in Fig.~\ref{fig3}. The presence of the CCPs does not affect this procedure since each of the upper and lower domains of $C_a$ gives a real-valued integrand in Eq.~(\ref{Eq:DoubleDiff}), which vanishes individually.
\par
As is conventionally discussed \cite{Chay:1990da, Bigi:1993fe, Manohar:1993qn, Blok:1993va}, the perturbative evaluation of the forward scattering tensor in the local OPE encounters an uncontrollable obstacle at the vicinity of the resonance region so that further deformation should be performed. Provided that the CCPs are absent, the contour integral along $C_a$ in Fig.~\ref{fig3} is related to the one for $C_b$ up to sign due to Cauchy's theorem. However, in the presence of the CCPs, the mentioned deformation leads to
\begin{eqnarray}
\frac{d^2\Gamma}{dE_\ell d\alpha}&=&
\left.\frac{d^2\Gamma}{dE_\ell d\alpha}\right|_{\rm pert}
+\left.\frac{d^2\Gamma}{dE_\ell d\alpha}\right|_{\rm CCPs}.\label{Eq:twoterms}
\end{eqnarray}
In Eq.~(\ref{Eq:twoterms}), the two terms read
\begin{eqnarray}
\left.\frac{d^2\Gamma}{dE_\ell d\alpha}\right|_{\rm pert}&=&
-\mathcal{F}(C_b, \tilde{T}),\label{Eq:double1}\\
\left.\frac{d^2\Gamma}{dE_\ell d\alpha}\right|_{\rm CCPs}
&=&\mathcal{F}(C_p, T_{\rm CCP})
+\mathcal{F}(C_{p^\prime}, T_{{\rm CCP}^\prime}),\label{Eq:double2}
\end{eqnarray}
where we defined
\begin{eqnarray}
\mathcal{F}(\mathcal{C}, \mathcal{T})
=\frac{G_F^2|V_{cb}|^2}{4\pi^3}\frac{m_b}{\pi}\textrm{Im}\int_{\cal C}(q^0-E_\ell)L_{\mu\nu}\mathcal{T}^{\mu\nu}dq^0.\nonumber
\end{eqnarray}
\par
Since $C_b$ is taken sufficiently away from the resonance region as shown in Fig.~\ref{fig3}, the perturbation theory gives a reliable prediction in Eq.~(\ref{Eq:double1}) so that we can replace $\tilde{T}^{\mu\nu}\to T^{\mu\nu}_{\rm pert}$. In what follows, the contribution of CCPs in Eq.~(\ref{Eq:double2}) is mainly considered since the perturbative contribution is discussed in the previous works \cite{Chay:1990da, Bigi:1993fe, Manohar:1993qn, Blok:1993va}.
\par
The integrals on the rhs of Eq.~(\ref{Eq:double2}) are evaluated straightforwardly by the residue theorem. Subsequently, the integral over $\alpha$ can be also performed, resulting in the lepton energy distribution from the one-pair CCPs
\begin{eqnarray}
      \left.\frac{1}{\Gamma_b}\frac{d\Gamma}{dy_\ell}\right|_{\rm CCPs}&=&
      G(R^{(1)}_0, Q_0),\label{eq:5}\\
      G(R, Q)&=&
      24\mathrm{Re}\left(R\left\{-\frac{1}{3}[1-(1-y_\ell)^{-3}]
      \right.\right.\nonumber\\
      &\times&(1-y_\ell+\hat{Q})+\left.\left.\frac{1}{2}[1-(1-y_\ell)^{-2}]\right.\right.\nonumber\\
&\times&\left.\left.(1+\hat{Q})\right\}\times(1-y_\ell+\hat{Q})^2\right).\label{eq:5prime}
\end{eqnarray}
In Eqs.~(\ref{eq:5}),(\ref{eq:5prime}), we introduced $\Gamma_b=G_F^2m_b^5|V_{cb}|^2/192\pi^3$, the dimensionless pole position denoted as $\hat{Q}=Q/m_b^2$, and the rescaled charged lepton energy, $y_\ell=2E_\ell/m_b$.
Moreover, some indices in Eq.~(\ref{eq:5}) are introduced as $R^{(1)}_0$ and $Q_0$ in order to distinguish them from the ones in the two-pair CCPs scenario similarly discussed later.
\par
It is worth noting that Eq.~(\ref{eq:5}) can be also derived from the partonic rate defined with $x_\ell = 1- \rho_c/(1-y_\ell)$
\begin{eqnarray}
\frac{1}{\Gamma_b}\frac{d\Gamma}{d y_\ell}=2y_\ell[3x_\ell^2 y_\ell (2-y_\ell)+x_\ell^3(y_\ell^2-3y_\ell)],\quad\label{Eq:parton}
\end{eqnarray}
by the replacement of $\rho_c=m_c^2/m_b^2\to -\hat{Q}_0$, multiplying $-2R^{(1)}_0$ as an overall factor, and taking the real part. This is interpreted as the relation between the pointlike on-shell condition for charm quark, dictated by $\delta(p_c^2-m_c^2)$, and the CCPs; the integrals for these two quantities fix $q^0$ to $(m_b^2-X+q^2)/2m_b$ with $X=m_c^2$ and $-Q^{(*)}$, leading to the mentioned correspondence.
\par
In the heavy quark limit, meson masses are approximated by the ones of quarks, $M_B\simeq m_b$ and $M_D\simeq m_c$. Integrating Eq.~(\Ref{eq:5}) with respect to $y_\ell$ for $0\le y_\ell\le 1-\rho_c$, one can obtain the total width from the one-pair CCPs
\begin{eqnarray}
    \frac{1}{\Gamma_b}\Gamma^{\mathrm{CCPs}}=\sum_{m=-3}^3
    c_mF_m,\label{eq:six}
\end{eqnarray}
where $F_m=(1-\rho^{m+1}_c)/(m+1)$ for $m\ne-1$, $F_{-1}=-\mathrm{log}(\rho_c)$, and
\begin{eqnarray}
c_3&=&-8\mathrm{Re}(R),\nonumber\\
c_2&=&12\mathrm{Re}[R(1-\hat{Q})],\nonumber\\
c_1&=&24\mathrm{Re}(R\hat{Q}),\nonumber\\
c_0&=&-4\mathrm{Re}[R(1+3\hat{Q}-3\hat{Q}^2-\hat{Q}^3)],\nonumber\\
c_{-1}&=&-24\mathrm{Re}(R\hat{Q}^2),\nonumber\\
c_{-2}&=&12\mathrm{Re}[R(1-\hat{Q})\hat{Q}^2],\nonumber\\
c_{-3}&=&8\mathrm{Re}(R\hat{Q}^3).\nonumber
\end{eqnarray}
\par
Equation~(\ref{eq:5}) can be generalized to the case with multiple pairs of CCPs by simply adding extra terms. The relevant expression for the scenario with two-pair CCPs is given by
\begin{eqnarray}
      \left.\frac{1}{\Gamma_b}\frac{d\Gamma}{dy_\ell}\right|_{\rm CCPs}&=&
      G(R_1^{(2)}, Q_1)+G(R_2^{(2)}, Q_2).
      \label{eq:TCCP}
\end{eqnarray}
Further generalization such as three-pair CCPs is rather straightforward. Likewise, one can obtain Eq.~(\ref{eq:six}) for the case with two-pair CCPs by integrating Eq.~(\ref{eq:TCCP}) in the phase space region.
\par
For $|V_{cb}|$, we consider the CCP corrections to the integrated semileptonic width in a way analogous to the discussion in Ref.~\cite{Crivellin:2014zpa},
\begin{eqnarray}
|V_{cb}|=\frac{|V_{cb}^{\mathrm{OPE}}|}{\sqrt{1+\frac{\tilde{\Gamma}^{\mathrm{CCPs}}}{\tilde{\Gamma}^{\mathrm{OPE}}}}}.\label{eq:seven}
\end{eqnarray}
In Eq.~(\ref{eq:seven}), we defined $\tilde{\Gamma}^{\rm CCPs}=\Gamma^{\rm CCPs}/|V_{cb}|^2$ and $\tilde{\Gamma}^{\mathrm{OPE}}\simeq G_F^2(m_{b}^\mathrm{kin})^5/192\pi^3$, where the latter with $m_{b}^\mathrm{kin}$ defined in the kinetic scheme~\cite{Bigi_1995} arises from the perturbative contour, while $|V_{cb}^{\rm OPE}|$ represents the CKM matrix element determined by the conventional method.
\par
\section{$B$-MESON LIFETIME}
Contributions of the CCPs from charm quarks can be also extracted for nonleptonic decays. We consider the total width of the $B_d^0$ meson, which consists of $b\to c\ell \bar{\nu}~(\ell = e, \mu)$, $b\to c\tau \bar{\nu}$, and $b\to c\bar{q}q^\prime~(q=u, c, q^\prime = d, s)$. It should be noted that transitions that proceed via $b\to u$ are negligible up to high accuracy due to the CKM suppression.
\par
The effective Hamiltonian relevant for $\Delta B=1$ nonleptonic decays reads \cite{Buchalla:1995vs}
\begin{eqnarray}
    \mathcal{H}_W=\frac{4G_F}{\sqrt{2}}V_{cb}V_{qq^\prime}^*(C_1O_1+C_2O_2),\label{eqWilson}
\end{eqnarray}
where the Wilson coefficients are denoted by $C_{i}~(i=1, 2)$ while the four-quark operators are
\begin{eqnarray}
O_1=(\bar{c}^\alpha\gamma_\mu P_Lb^\alpha)(\bar{q}^{\prime\beta}\gamma^\mu P_Lq^\beta),\nonumber\\ 
O_2=(\bar{c}^\alpha\gamma_\mu P_Lb^\beta)(\bar{q}^{\prime\beta}\gamma^\mu P_Lq^\alpha).\nonumber
\end{eqnarray}
In the above relations, $\alpha$ and $\beta$ represent color indices.
\par
\begin{figure}[H] 
	\centering  
	\vspace{-0.35cm} 
	\subfigtopskip=2pt 
	\subfigbottomskip=2pt 
	\subfigcapskip=-2pt 
	\subfigure[\text{$\theta_{i}=\pi/2,~n=8$}]{
		\label{sub1}
        
		\includegraphics[width=0.45\linewidth]{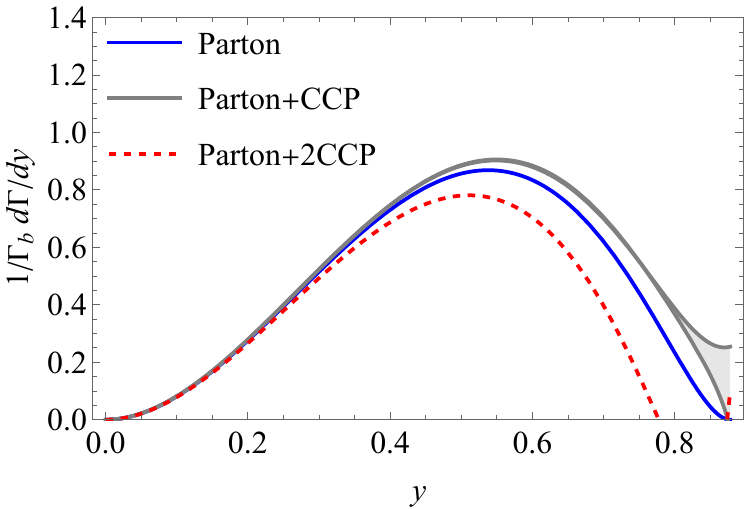}}
	\quad 
	\subfigure[$\theta_{i}=\pi/2,~n=4$]{
		\label{sub2}
		\includegraphics[width=0.45\linewidth]{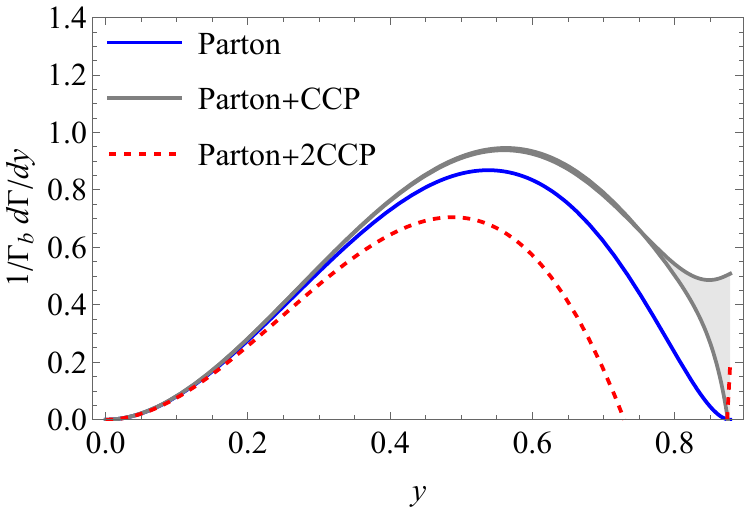}}
	\subfigure[$\theta_{i}=\pi,~n=8$]{
		\label{sub3}
		\includegraphics[width=0.45\linewidth]{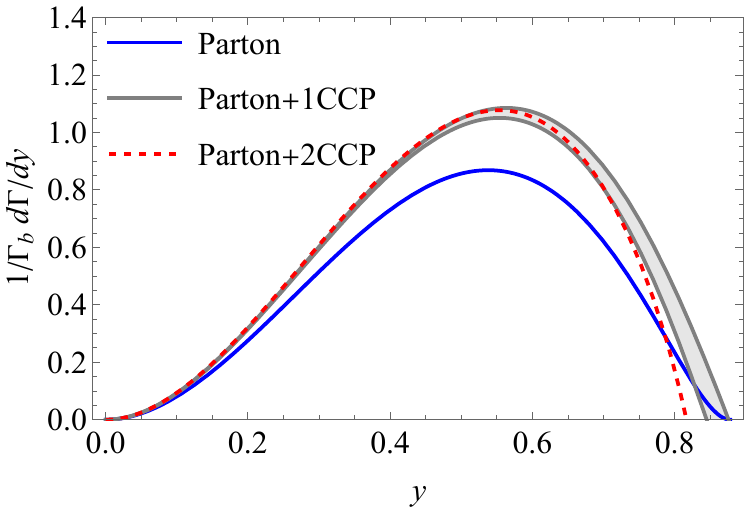}}
	\quad
	\subfigure[$\theta_{i}=\pi,~n=4$]{
		\label{sub4}
		\includegraphics[width=0.45\linewidth]{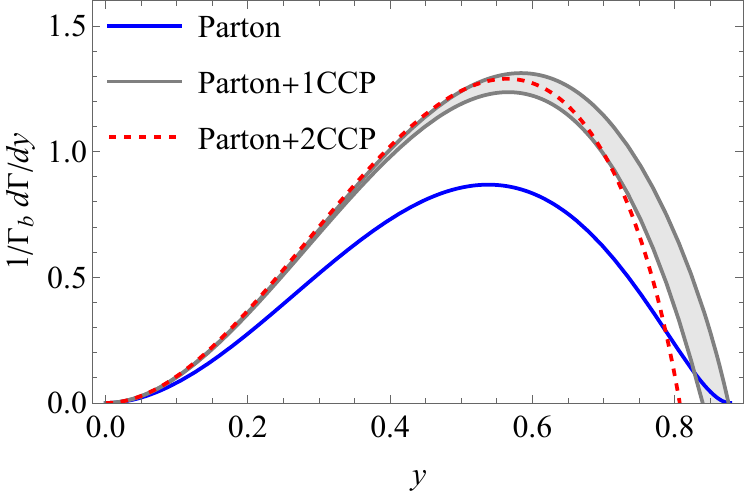}}

        \subfigure[$n=8$]{
		\label{sub5}
		\includegraphics[width=0.45\linewidth]{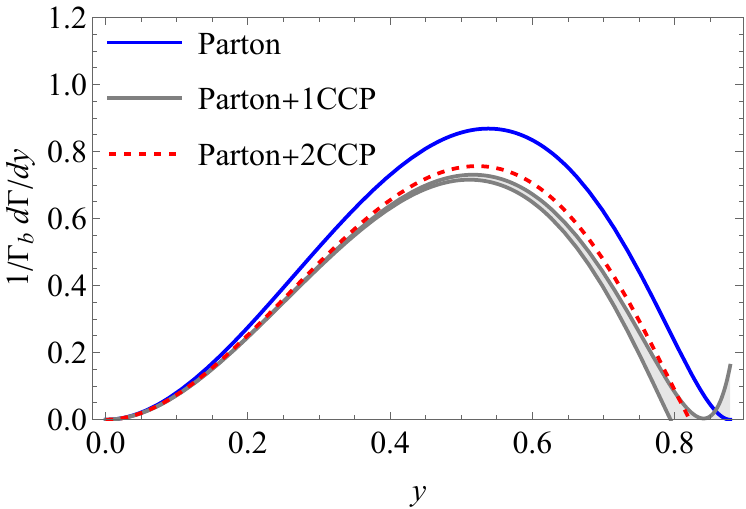}}
	\caption{\label{fig4}Lepton energy distribution for $B\to X_c \ell\bar{\nu}$ decay. The blue solid line represents the parton-level contribution while the gray band (red dotted line) shows the one additionally including the contributions of one-pair (two-pair) CCPs. The gray band region represents the uncertainty range of $\textrm{Re}~Q_0$ in Ref.~\cite{Zhu:2020bwu}. In each plot, the values of $\theta_i$ (i=0,1,2) are common.}
	
\end{figure}
In the Born approximation, widths of nonleptonic \textit{B} decays differ from leptonic cases up to overall factors of $\tilde{C}=N_cC_1^2+N_cC_2^2+2C_1 C_2$ and a CKM matrix element: Relations between the widths in the massless limits for $u, d, s, e, \mu$ and neutrinos are
\begin{eqnarray}
    \frac{d\Gamma^{b\to c\bar{u}q^\prime}}{dy_{q\prime}}&=& |V_{uq^\prime}|^2\tilde{C}\left.\frac{d\Gamma^{b\to c\ell \bar{\nu} }}{dy_\ell }\right|_{y_\ell\to y_{q^\prime}},\qquad \quad
    \label{eqfactor}\\
    \frac{d\Gamma^{b\to c\bar{c}q^\prime}}{dy_{\bar{c}}}&=& |V_{cq^\prime}|^2\tilde{C}\left.\frac{d\Gamma^{b\to c\tau\bar{\nu} }}{dy_\tau }\right|_{y_\tau\to y_{\bar{c}},~\rho_{\tau}\to \rho_{\bar{c}}},\qquad \quad
    \label{eqfactor2}
\end{eqnarray}
where we defined $y_i=2E_i/m_b$ with $i$ being an appropriate flavor, $\rho_{\tau(\bar{c})} = [m_{\tau (c)}/m_b]^2$, and $x_{\tau}=1-\rho_{\bar{c}}/(1+\rho_\tau-y_\tau)$. In Eq.~(\ref{eqfactor}), the massless semileptonic decay rate is defined in Eq.~(\ref{Eq:parton}) while the case with $\tau$ in Eq.~(\ref{eqfactor2}) is \cite{Falk:1994gw}
\begin{eqnarray}
\frac{1}{\Gamma_b}\frac{d\Gamma^{b\to c\tau \bar{\nu}}}{dy_\tau}&=&
2\sqrt{y_\tau^2-4\rho_\tau}\{
3x_\tau^2(y_\tau-2\rho_\tau)(2-y_\tau)\nonumber\\
&&+x_\tau^3[y_\tau^2-3y_\tau(1+\rho_\tau)+8\rho_\tau]
\}.
\end{eqnarray}
\par
In what follows, the one pair of CCPs is considered for definitiveness, while the multiple-pair case is discussed in the later straightforward generalization. The relations between nonleptonic and semileptonic modes in Eqs.~(\ref{eqfactor}),(\ref{eqfactor2}) are valid also for the CCP contributions. By utilizing these relations and the correspondence between the partonic rate and the one from CCPs, discussed around Eq.~(\ref{Eq:parton}), we can obtain CCP contributions to the nonleptonic widths; the results are ones on the lhs~in Eqs.~(\ref{eqfactor}),(\ref{eqfactor2}) with the replacement of $\rho_{c}\to -\hat{Q}$, multiplying $-2R$ as an overall factor, and taking the real part. These can be also extracted from the direct calculation by the residue theorem. Integrating the lhs~in Eqs.~(\ref{eqfactor}),(\ref{eqfactor2}) for $0\leq y_{q'}\leq 1-\rho_c$ and $2\sqrt{\rho_{\bar{c}}}\leq y_{\bar{c}}\leq 1$, respectively, the nonleptonic widths are evaluated.
\par
For the processes including $b\to c\bar{c}q^\prime$, CCPs from the $\bar{c}$ quark contribute in addition to ones from the $c$ quark. This can be obtained by proper procedure with Eq.~(\ref{eqfactor2}) described as follows: after interchanging $c\leftrightarrow \bar{c}$, one carries out the replacement of $\rho_{\bar{c}}\to -\hat{Q}$, multiplies $-2R$, and takes the real part as was done before. Implemented in this way, one can find that the CCP contributions from $\bar{c}$ to the integrated width are identical to the ones from $c$. In what follows, these two types of contributions are both included for evaluating the widths of $b\to c\bar{c}q^\prime$ decays.
\par
By summing over the final states, including nonleptonic and semileptonic modes, one can obtain the CCP contributions from charm quark to \textit{B}-meson lifetime
\begin{eqnarray}
\tau(B_d^0)=\frac{1}{\Gamma_{\mathrm{all}}^{\mathrm{CCPs}}+\Gamma^{\mathrm{OPE}}},\label{eq:eight}
\end{eqnarray}
where $\Gamma^{\mathrm{OPE}}$ is the OPE contribution while $\Gamma_{\mathrm{all}}^{\mathrm{CCPs}}$ is that from the CCPs. For the multiple-pair CCPs, extra terms should be added in $\Gamma_{\mathrm{all}}^{\mathrm{CCPs}}$, as was discussed for semileptonic decays in Eq.~(\ref{eq:TCCP}).
\par
\section{NUMERICAL RESULT}
For the input parameters, the kinetic mass and the pole masses are respectively set to $m_b^{\rm kin}=4.573~\mathrm{GeV}$ \cite{Bordone:2021oof}, $m_b=4.78~\mathrm{GeV}$ \cite{ParticleDataGroup:2024cfk}, and $m_c=1.67~\mathrm{GeV}$ \cite{ParticleDataGroup:2024cfk}.\par
In what follows, we investigate two scenarios, where the number of pairs of CCPs is one or two, respectively. The pole positions and residues of the charm quark propagator are extracted for the former \cite{Zhu:2020bwu} and the latter \cite{Dorkin:2013rsa} cases. It should be noted that the size of the CCP contributions is characterized by the residue, since this serves as an overall factor of the nonanalytical term. Provided that the CCP contributions give a slight correction to the partonic rate, an immediate observation is that the typical size of the residues in the previous works \cite{Zhu:2020bwu, Dorkin:2013rsa} is too large. In view of phenomenological relevance, the residues in Eqs.~(\ref{eq:5}),(\ref{eq:TCCP}) are represented by
\begin{eqnarray}
    R_0^{(1)}=\frac{|R_0|}{n}e^{i\theta_0},\qquad
    R_{1, 2}^{(2)}=\frac{|R_{1, 2}|}{n}
    e^{i\theta_{1, 2}},\label{Rparam}
\end{eqnarray}
with $n$ being $8, 4$ or $3$. $|R_0|$ and $|R_{1, 2}|$ are fixed as absolute values from,
\begin{eqnarray}
    R_0&=&0.577 - 0.712i~[38],\label{Eq:R0val}\\
    R_1&=&0.08624 + 0.598i~[53],\label{Eq:R1val}\\
    R_2&=&0.4145 - 0.267i~[53],\label{Eq:R2val}
\end{eqnarray}
while for the pole positions, we directly adopt values obtained in the previous works,
\begin{eqnarray}
    Q_0&=&(-2.325+1.145i)~\textrm{GeV}^2~[38],\\
    Q_1&=&-(1.773+0.7179i)^2~\textrm{GeV}^2~[53],\\
    Q_2&=&-(2.112+0.5177i)^2~\textrm{GeV}^2~[53].
\end{eqnarray}
Furthermore, in the scenario with one-pair CCPs, we consider the uncertainty range of $-3\leq \textrm{Re}~Q_0/\textrm{GeV}^2\leq-2$ inferred from Fig.~4 in Ref.~\cite{Zhu:2020bwu}, which covers the mentioned value, $\textrm{Re}~Q_0=-2.325~\textrm{GeV}^2$.
\par
In Fig.~\ref{fig4}, the lepton energy distributions for $B\to X_c\ell\bar{\nu}$ decays based on Eqs.~(\ref{eq:5}),(\ref{Eq:parton}),(\ref{eq:TCCP}) are displayed for illustrating the size of the CCP contributions. For (a)-(d), the results for the parametrization in Eq.~(\ref{Rparam}) with $\theta_i=\pi/2, \pi~(i=0, 1, 2)$ and $n=8, 4$ are respectively exhibited. The uncertainty ranges from $\textrm{Re}~Q_0$, mentioned before, are represented as gray bands. In (e), $n$ is fixed to $8$ while $\theta_i$s are extracted as arguments of the residues in Eqs.~(\ref{Eq:R0val})-(\ref{Eq:R2val}). As one can see from the results, how large the CCPs contributions are depends crucially on the parameters for residues and pole positions. The case with $n=4$ is interpreted as the scale-up of the CCP contributions for $n=8$. Taking Fig.~\ref{fig4} (c) and (d) as examples, for $y=0.5$, the one-pair (two-pair) scenario gives approximately
$21\%$ ($22\%$) for $n=8$ and $43\%$ ($45\%$) for $n=4$  larger results than those of the partonic rate. If we take $0<n<4$, the two scenarios give further large corrections. For (c)-(e), one can find that the results for the one-pair and two-pair scenarios are rather close to each other except for large $y$, while for (a)-(b), the results are deviated for smaller $y$.
\begin{figure}[t]
\includegraphics[width=8cm,height=5.5cm]{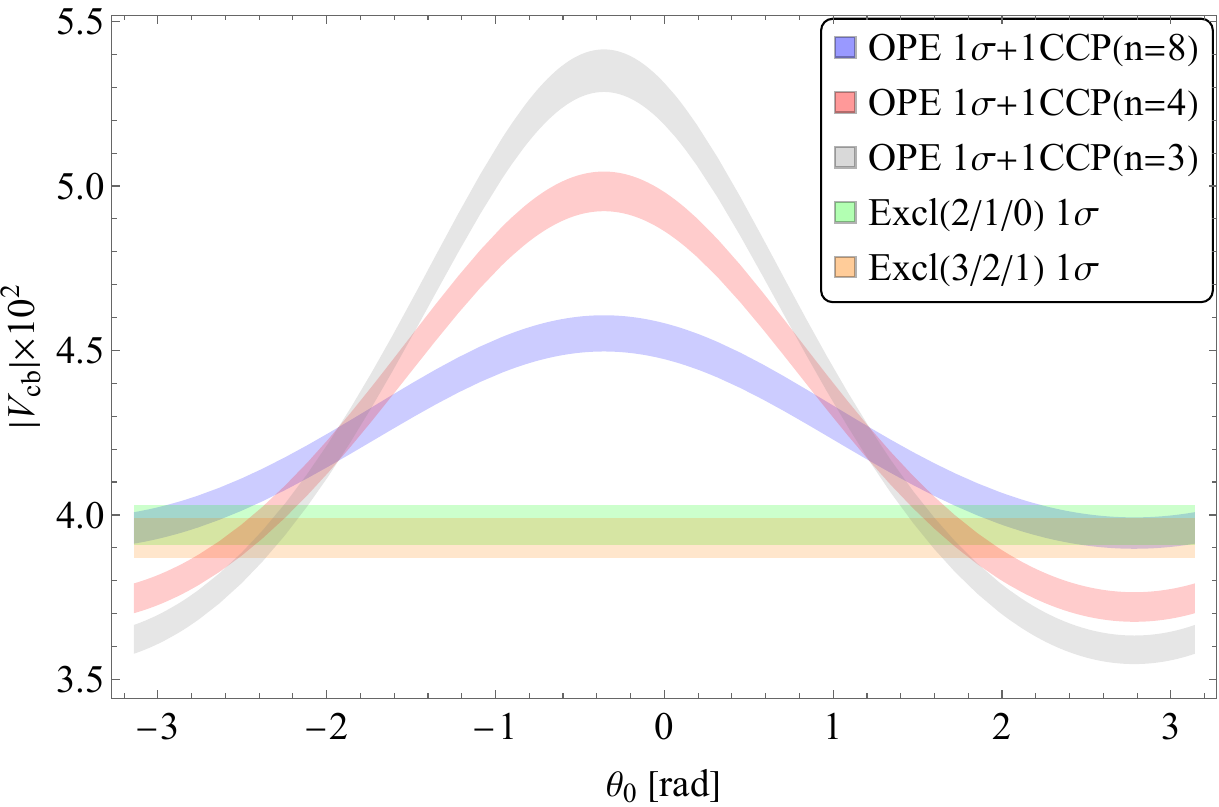}
\caption{\label{fig1} Dependence of $|V_{cb}|$ on the argument of the residue parameter in the scenario with one-pair CCPs. The blue, red and gray bands represent the results determined by the OPE term plus the CCPs for $n=8, 4$ and $3$, with the 1$\sigma$ uncertainty range from $|V_{cb}^{\rm OPE}|$ \cite{Bordone:2021oof}. The green and orange bands are the $1\sigma$ range of the exclusive determinations \cite{Iguro:2020cpg} in the fitting scenarios of $(2/1/0)$ and $(3/2/1)$, respectively.}
\end{figure}
\begin{figure}[t]
\includegraphics[width=8cm,height=5.5cm]{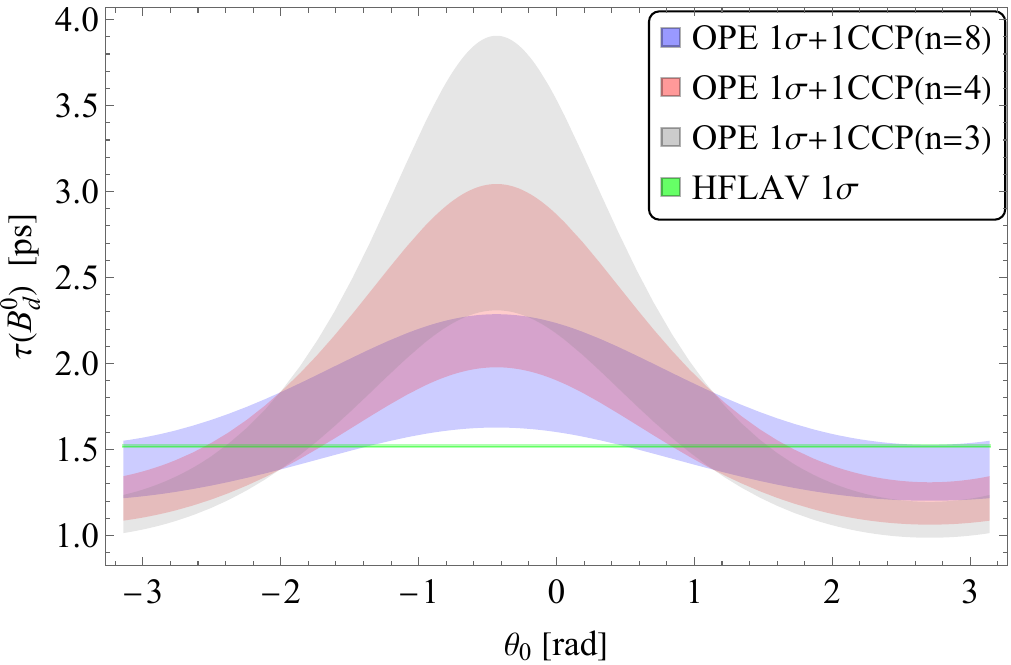}
\caption{\label{fig2} Dependence of $B_d^0$ lifetime on the argument of the residue parameter in the scenario with one-pair CCPs. The blue, red, and gray bands represent the results including both the OPE \cite{Lenz:2022rbq} and the CCP contributions for $n=8, 4$ and $3$, where the $1\sigma$ uncertainty range from the OPE \cite{Lenz:2022rbq} is displayed. The green line shows the experimental data from Heavy Flavor Averaging Group (HFLAV) \cite{HFLAV:2022esi}.}
\end{figure}
\par
For the analysis of $|V_{cb}|$, a recent inclusive determination for $|V_{cb}^{\mathrm{OPE}}|$ is adopted in Eq.~(\Ref{eq:seven}). As to exclusive determinations that are compared with this work, there are results based on parametrizations of form factors from Boyd-Grinstein-Lebed \cite{Boyd:1995cf}, Caprini-Lellouch-Neubert \cite{Caprini:1997mu}, one relying on the HQET \cite{Jung:2018lfu, Iguro:2020cpg}, etc. We adopt the third formalism, since analyticity is not intrinsically assumed for this parametrization. Those input parameters are summarized below
\begin{eqnarray}
    |V_{cb}^{\rm OPE}|&=&(42.16\pm0.50)\times 10^{-3}\label{eq:one}~[3], \nonumber\\
    |V_{cb}^{\rm exc}|&=&(39.7\pm 0.6)\times 10^{-3}~\quad\text{(2/1/0)}~[54],\nonumber\\
    |V_{cb}^{\rm exc}|&=&(39.3\pm 0.6)\times 10^{-3}~\quad\text{(3/2/1)}~[54],\nonumber
\end{eqnarray}
where the last two results correspond to exclusive fitting scenarios that include different powers in the recoil-variable expansion.
\par
For the inputs to evaluate $\tau(B_d^0)$ in Eq.~(\ref{eq:eight}), the conventional OPE result is given by $\Gamma^{\mathrm{OPE}}=(0.615^{+0.108}_{-0.069})$~$\mathrm{ps^{-1}}$ \cite{Lenz:2022rbq}, predicted by the parameter set in Ref.~\cite{Bordone:2021oof}. The Wilson coefficients are given at next-to-leading order in QCD corrections, $C_1(m_b)=1.07$ and $C_2(m_b)=-0.17$ based on Ref.~\cite{Buchalla:1995vs}. The CKM matrix elements for $|V_{ij}|~(i=u, c, j=d, s)$ and exclusive $|V_{cb}|=39.8\times 10^{-3}$ are extracted from PDG~\cite{ParticleDataGroup:2024cfk}.
\par
In Fig.~\ref{fig1}, inclusive $|V_{cb}|$ for the scenario with one-pair CCPs in Eq.~(\Ref{eq:seven}) is exhibited as a function of $\theta_0$ for $n=8, 4$ and $3$, and compared with the exclusive results. One can find that the inclusive result for $n=8$ around $\theta_0=\pm\pi$ is consistent with the exclusive ones within $1\sigma$. Moreover, $B_d^0$-meson lifetime including the CCP contributions is displayed, and compared with the experimental data \cite{HFLAV:2022esi} in Fig.~\ref{fig2}. It should be noted that the theoretical uncertainty of lifetimes is much larger than that for the experimental data. One can find from Fig.~\ref{fig2} that the theoretical $\tau(B_d^0)$ for $n=8$ is consistent with the experimental data within 1$\sigma$ in a neighborhood of $\theta_0=\pm\pi$, similar to Fig.~\Ref{fig1} with the same value of $n$. Hence, there exists a parameter region for $n=8$ where $|V_{cb}|$ and $\tau(B_d^0)$ are simultaneously explained within the $1\sigma$ range. As to the cases of $n=4~(3)$, we similarly find that $\theta_0\approx -2.4~\textrm{rad}~(-2.3~\textrm{rad})$ and $\theta_0\approx 1.6~\textrm{rad}~(1.5~\textrm{rad})$ simultaneously explain the $|V_{cb}|$ and $\tau(B_d^0)$ within $1\sigma$. 
\par
\begin{figure}[t] 
	\centering  
	\vspace{-0.35cm} 
	\subfigtopskip=2pt 
	\subfigbottomskip=2pt 
	\subfigcapskip=-2pt 
	\subfigure[$n=8$]{
		\label{sub1}
		\includegraphics[width=0.381\linewidth]{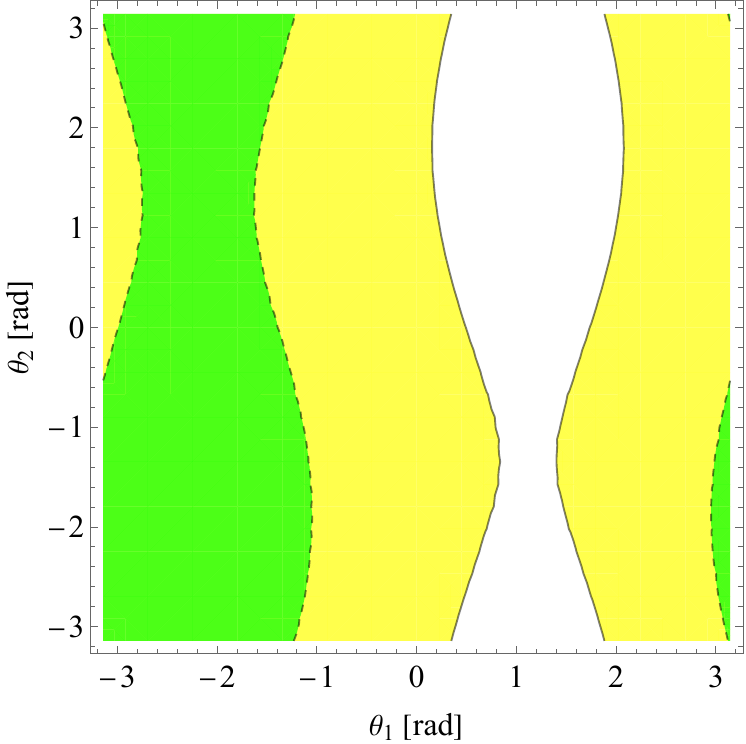}}
	\quad 
	\subfigure[$n=4$]{
		\label{sub2}
		\includegraphics[width=0.538\linewidth]{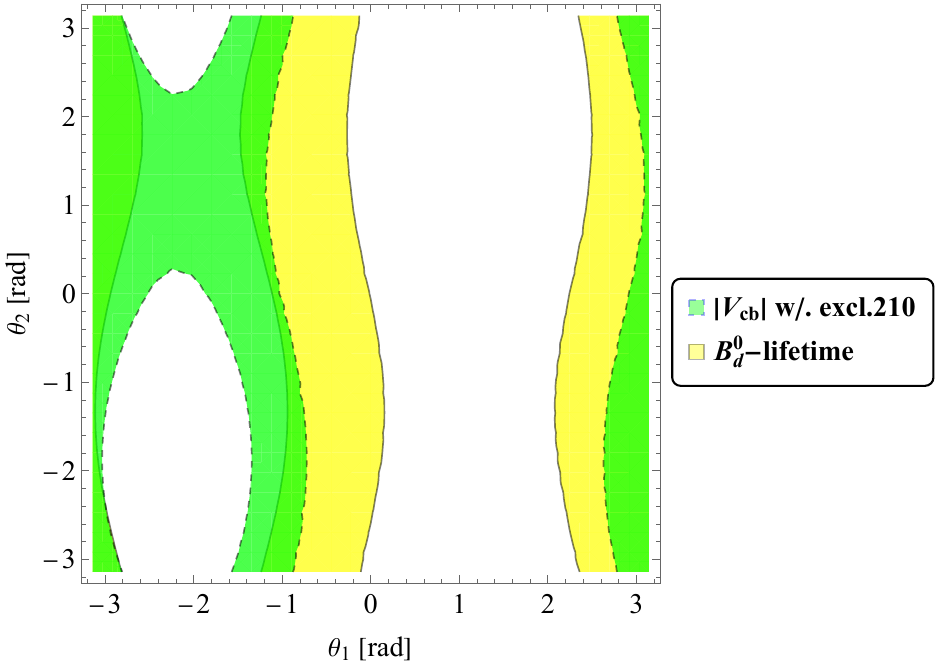}}
        \caption{\label{fig5} Allowed parameter regions in the scenario with two-pair CCPs. The green band represents the $1\sigma$ region for inclusive $|V_{cb}|$ consistent with the exclusive scenario (2/1/0) \cite{Iguro:2020cpg}. The yellow band represents the $1\sigma $ region for the $B_d^0$-meson lifetime consistent with the HFLAV data \cite{HFLAV:2022esi}.}
\end{figure}

In the scenario with two-pair CCPs, the results are displayed in Fig.~\ref{fig5}. For definitiveness, the (2/1/0) fitting of the exclusive $|V_{cb}|$ \cite{Iguro:2020cpg} is considered in the figure. As can be seen from the results, there are regions that simultaneously explain $|V_{cb}|$ and $\tau(B_d^0)$ within $1\sigma$ for both $n=8$ and $4$, as well as the case with the one-pair CCPs. It should be noted that the allowed region for $\theta_1$ and $\theta_2$ is larger for $n=8$ than that for $n=4$ since in the latter case the typical size of the CCP contributions is larger, and therefore limiting the available parameter space.
\par
\section{CONCLUSION}
In this work, we have discussed the phenomenological aspects of the nontrivial contributions in the quark propagator. It should be remembered that the nonanalytical structures are not definitively concluded in the previous works, especially for pole positions, the residue parameters, and the number of pairs of the CCPs. In view of these complexities, we have carried out illustrative numerical investigations for $B$-meson inclusive decays.
\par
In the presence of the CCPs, deformation of the integration contour, implemented for avoiding the resonance region, leads to the nonperturbative corrections in addition to the OPE contributions. This gives rise to the limitation of the accuracy in the OPE analysis even if the smearing over phase space is performed sufficiently. We have extracted the mentioned contribution by utilizing the residue theorem, and found that the result is also obtainable from a straightforward manner where charm quark mass square in the partonic rate is replaced by the complex-valued one, with a proper modification of the overall coefficient.
\par
As to phenomenological observables, the integrated rate for $B\to X_c\ell\bar{\nu}$ and $B_d^0$-meson lifetime are analyzed. In order to ensure that the CCP contributions give a small correction to the partonic rate, which is of phenomenological relevance, it is found that the residue parameter must be somewhat smaller than those in Refs.~\cite{Zhu:2020bwu, Dorkin:2013rsa}, unless one tunes the parameters for CCPs. Viewed from another way, the absolute value of the residue must be sufficiently suppressed in such a way that the CCPs do not give too large corrections to observables.
\par
We have performed the numerical analysis by varying the parameters of the CCPs. Some illustrative cases, where the size of the residue parameters is smaller than ones in Refs.~\cite{Zhu:2020bwu, Dorkin:2013rsa} and the number of the pairs of CCPs is either one or two, are tested. It is shown that there exist parameter regions accommodating $|V_{cb}|$ and $\tau(B_d^0)$ within $1\sigma$. Further phenomenological clarification requires, not just discussing the integrated semileptonic rate, fitting analysis for, e.g., the lepton energy moments as well as $q^2$ moments \cite{Bordone:2021oof, Bernlochner:2022ucr}, to extract $|V_{cb}|$.
\par
The result in this work leads to specific modeling of duality violation, in addition to the previous ones from the instanton-based and resonance-based approaches. Here \enquote{modeling} is phrased in the sense that nontrivial aspects of gauge theories are involved; how many pairs of CCPs exist is not even clearly identified, see, e.g., \cite{Zhu:2020bwu, Dorkin:2013rsa}. The phenomenological discussion in $B$-meson decays enables us to test such an aspect of gauge theories.
\par
Duality violation in the present work is to be contrasted with ones in the previous discussions, which have some drawbacks: Even if one starts from the instanton-based approach, realistic size of duality violation is still based on another modeling such as the single-instanton approximation, adopted in Refs.~\cite{Chay:1994si,Chay:1994dk,Falk:1995yc,Chibisov:1996wf,Umeeda:2022dpt}, or more preferably the instanton-liquid model \cite{Shuryak:1981ff}. As for the 't Hooft model, formulated in two-dimensional spacetime and large-$N_c$ limit, one cannot obtain an obvious insight on how the result is altered in four dimensions with $N_c=3$. Meanwhile, if one fixes the parameters for CCPs, it rather simply leads to duality violation. In this sense, the model based on CCPs does not involve technical complexities, once the residue and the pole position are given.
\par
Due to the difficulty to quantify quark-hadron duality, one cannot give a definitive conclusion on whether the $|V_{cb}|$ puzzle is caused by CCPs or others. On the other hand, the primary finding in this work is that nontrivial analyticity, still under active debate, practically alters the inclusive rate. If the parameters for CCPs are determined by the research directions in, e.g., Refs.~\cite{Zhu:2020bwu, Dorkin:2013rsa}, the quantitative {\it prediction} can be obtained for $B$-meson decays, without entailing \textit{ad hoc} choices of the parameters. Obviously, further phenomenological discussion of the CCPs, not limited to heavy quark physics, is required so as to clarify novel nonanalytical structures of gauge theories.
\par
\section*{Acknowledgment}
We gratefully thank Gael Finauri for the comment. This work is supported by the National Science Foundation of China (NSFC) under Grant No.~12405111 and the Seeds Funding of Jilin University. 

\bibliography{arXiv_v1}

\end{document}